\documentclass[prd,aps,showpacs,secnumarabic]{revtex4}
\usepackage[dvips]{graphicx,color}
\usepackage{amssymb,amsmath,bm}
\def\ga{\gamma}

\begin{document}
\title{Calculation of the decay ${\bm H\bm\to\bm e\bar{\bm e}\bm\ga}$}
\author{Duane A. Dicus} \email{dicus@physics.utexas.edu}\affiliation{Center for Particle Physics, University of Texas, Austin, TX 78712, USA} 
\author{Wayne W. Repko}\email{repko@pa.msu.edu} \affiliation{Department of Physics and Astronomy, Michigan State University, East Lansing, MI 48824, USA}
\date{\today}
\begin{abstract}
We revisit an earlier calculation of the decay $H\to e\,\bar{e}\ga$ using the recently reported mass value of the Higgs boson candidate observed in the ATLAS and CMS experiments together with cuts that are appropriate for experimental analyzes.
\end{abstract}
\maketitle
After the announcements of a candidate for the Standard Model Higgs boson by the ATLAS and CMS experiments \cite{ATLAS,CMS}, there has been an intense effort to determine the precise couplings on this particle to the particles of the Standard Model. The most accessible decay modes have been explored by each of these experiments. In the fermion sector, the $H\to\tau\bar{\tau}$ and $H\to b\bar{b}$ decays have been measured as have the $H\to ZZ^*$, $H\to WW^*$ and $H\to\ga\ga$ in the gauge boson sector.

As more data is accumulated, rarer decays will become accessible. Among these is the decay $H\to e\,\bar{e}\ga$, which we were involved in calculating in the pre-LHC days \cite{ABCDR}. Since the Standard Model direct coupling of an $e\,\bar{e}$ to a Higgs boson is negligibly small, this decay occurs, for all intents and purposes, at the one-loop level and consequently involves Standard Model couplings to both gauge bosons and heavy quarks. In this report, we revisit our original calculation by using a Higgs boson mass of 125 GeV and imposing cuts on the decay products that are consistent with those commonly used in LHC analyzes.

\begin{figure}[h]\centering
\includegraphics[height=2.0 in]{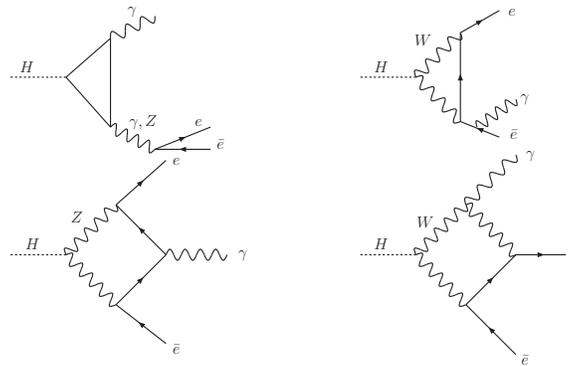}
\caption{Typical triangle and box diagrams for \protect{$H\to e\,\bar{e}\gamma$} are shown. \label{diag}}
\end{figure}

In Ref.\cite{ABCDR}, we computed all the relevant one-loop diagrams in an $R_\xi$ gauge using a non-linear gauge fixing term for the $W$ boson field. Use of this non-linear gauge eliminates the $W$ boson - charged Goldstone boson - photon coupling that occurs in the usual formulation, reducing the number of diagrams considerably and making the triangle and box diagrams separately gauge invariant. All the invariant amplitudes are found in the Appendices of Ref.\cite{ABCDR}. Typical diagrams are shown in Fig.\,\ref{diag}.
In evaluating these amplitudes, we have used the physical (non-zero) value for the electron mass.  However, because Ref.\,\cite{ABCDR} provides exact analytic expressions for all the amplitudes \cite{ABCDR_1}, it is easy to see that any potentially singular dependence on the electron mass cancels in the sum of the amplitudes - there are no $\ln(m_e)$ terms.  The electron mass does appear in the invariant masses and the phase space.

\begin{figure}[h]
\includegraphics[height=3.0 in]{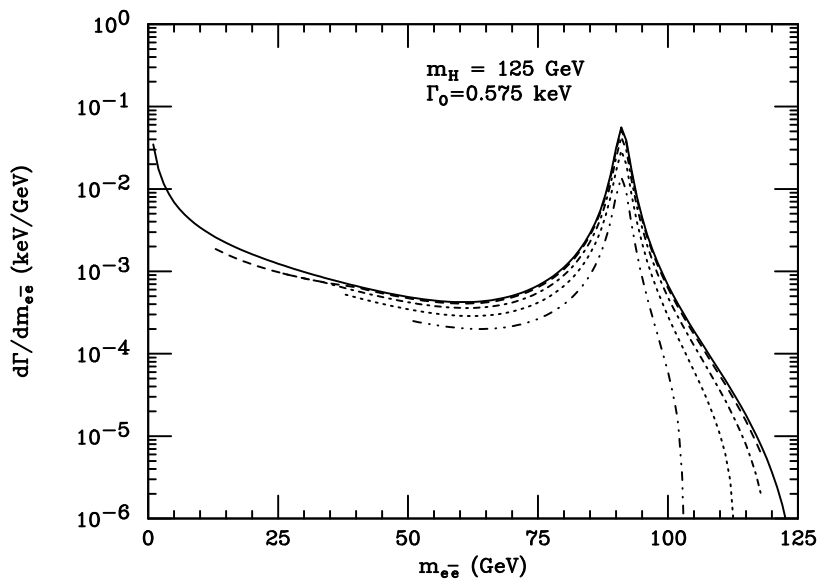}
\caption{The solid line has no cuts and the remaining curves have the lepton and photon energy cuts mentioned in the text and the invariant mass cuts of Eq.\,(\ref{mcuts}). The dashed line corresponds to $k=0.1$, the dot-dashed line to $k=0.2$, the dotted line to $k=0.3$ and the dot dot dashed line to $k=0.4$. \label{cuts}}
\end{figure}

Now that the Higgs boson candidate is known to have a mass in the vicinity of 125 GeV, its Standard Model decay width into $e\,\bar{e}\ga$ can be computed. More importantly, the experimental cuts usually applied in the analysis of electron and photon events can be implemented to see how the decay width changes. For our calculation, the cuts we imposed were that one lepton had an energy greater than 25 GeV, the other had an energy greater than 7 GeV and the photon had an energy greater than 5 GeV. In addition, we assumed that the invariant masses $m_{e\bar{e}}$, $m_{e\ga}$ and $m_{\bar{e}\ga}$ each satisfied 
\begin{equation}\label{mcuts}
m^2_{e\bar{e}}=m^2_{e\ga}=m^2_{\bar{e}\ga}\geq (k\,m_H)^2\,,
\end{equation}
with $k=0.1,\,0.2,\,0.3$, or $0.4$. The $m_{e\bar{e}}$ invariant mass distributions with these cuts are shown in Fig.\,\ref{cuts}. The importance of the $Z$ pole is clearly shown. If there are no cuts, the photon pole is also very important because the minimum value of $m_{e\bar{e}}$ is twice the electron mass.

The corresponding widths are obtained by integrating the $m_{e\bar{e}}$ invariant mass distributions. The variations of the widths (in keV) for the range 123 GeV $\leq m_H\leq$ 127 GeV are given by
\begin{eqnarray}
\Gamma_0 &=& 0.575 + 0.032\,(m_H - 125\,{\rm GeV}) \label{gamvar0} \\
\Gamma_{0.1} &=& 0.243 + 0.020\,(m_H - 125\,{\rm GeV})\label{gamvar1} \\
\Gamma_{0.2} &=& 0.188 + 0.017\,(m_H - 125\,{\rm GeV})\label{gamvar2} \\
\Gamma_{0.3} &=& 0.123 + 0.012\,(m_H - 125\,{\rm GeV})\label{gamvar3} \\
\Gamma_{0.4} &=& 0.0577 + 0.0074\,(m_H - 125\,{\rm GeV})\label{gamvar4}\,.
\end{eqnarray}
Here, $\Gamma_0$ is the full width while the others include the cuts with the particular $k$ values. All figures use $m_H=125$ GeV. Imposition of the $k=0.1$ cut reduces the width by a factor of about two, essentially eliminating the photon pole enhancement. The larger values of $k$ further reduce the width, though less drastically. 

The width with the cuts $m^2_{e\ga}=m^2_{\bar{e}\ga}=(0.1m_H)^2$ and $m^2_{e\bar{e}}=(0.6m_H)^2$, introduced to emphasize the $Z$ pole region, varies over the same Higgs boson mass range as
\begin{equation}
\Gamma = 0.198 + 0.018\,(m_H - 125\,{\rm GeV})\,.
\end{equation}
This case is not shown in any of the figures.

\begin{figure}[h]
\includegraphics[height=3.0 in]{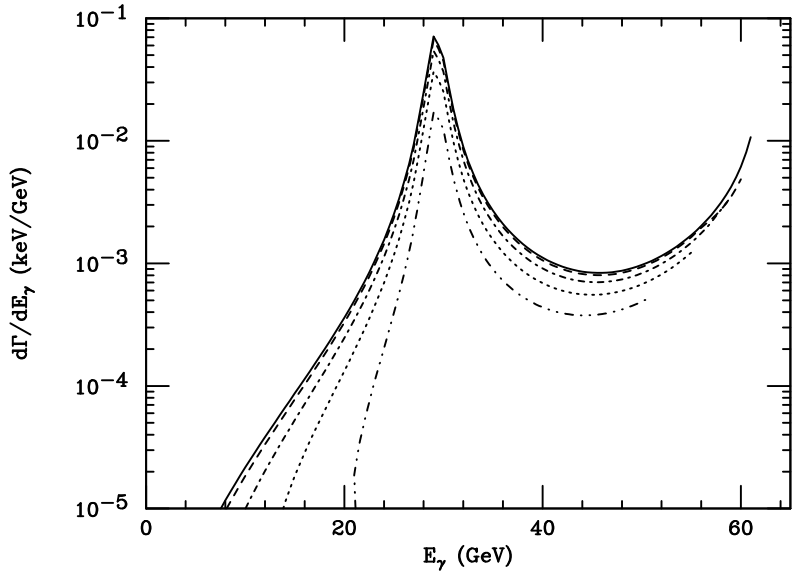}
\caption{The invariant distribution of the photon energy $E_\ga$ for the decay $H\to e\bar{e}\ga$ is shown with the same sets of cuts used in the $m_{e\bar{e}}$ distributions. \label{gdist_ee}}
\end{figure}

It may be of interest to consider the distribution of the photon energies in the $H\to e\bar{e}\ga$ decay and these are shown in Fig.\,\ref{gdist_ee} with the same cuts used in the $m_{e\bar{e}}$ distributions. In the Higgs rest frame, $E_\ga=(m_H^2-m_{e\bar{e}}^2)/2m_H$.

\begin{figure}[h]
\includegraphics[height=3.0 in]{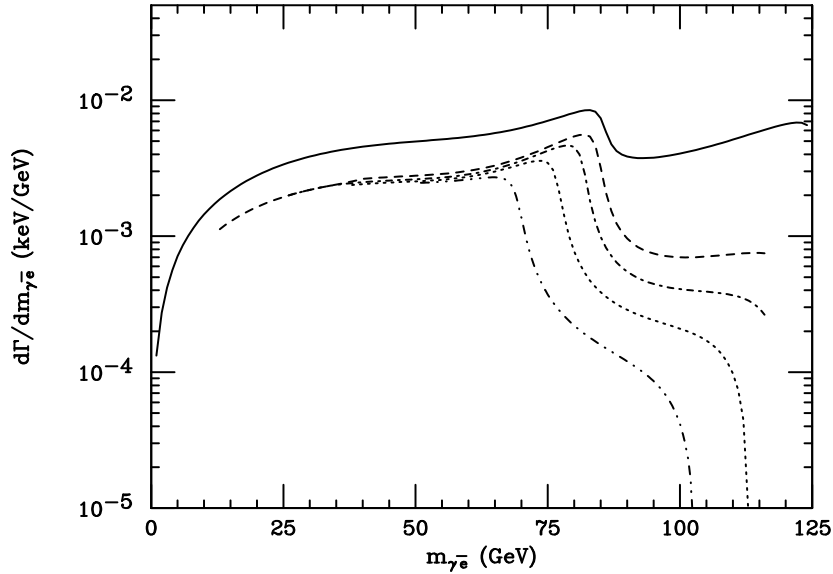}
\caption{The $m_{\ga\bar{e}}$ invariant distributions are shown with the same sets of cuts used in the $m_{e\bar{e}}$ distributions. \label{Heeg_125_dGdu}}
\end{figure}

Using the same cuts, we can also calculate the $m_{\ga\bar{e}}$ (or $m_{\ga e}=m_{\ga\bar{e}}$) invariant mass distributions and these are shown in Fig.\,\ref{Heeg_125_dGdu}. The solid line denotes the distribution with no cuts. When integrated over $m_{\ga\bar{e}}$, the curves reproduce the results in Eqs.\,(\ref{gamvar0}-\ref{gamvar4}) for $m_H=125$ GeV. This distribution involves an integral over $m_{e\bar{e}}$ and the sharp breaks at $70-85$ GeV result because the range of that integration no longer includes the $Z$ pole. 

\begin{figure}[h]
\includegraphics[height=3.0 in]{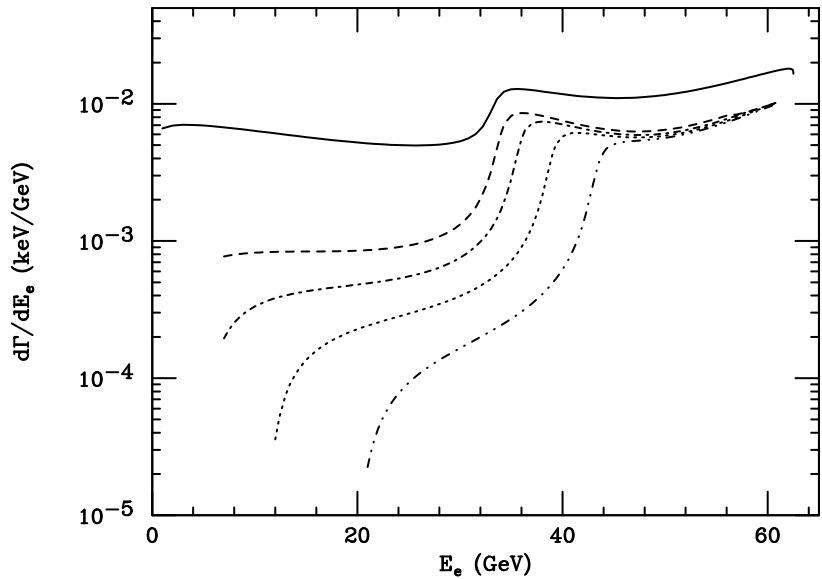}
\caption{The distributions of the electron energies are shown with the same sets of cuts used in the $m_{e\bar{e}}$ distributions. \label{dGdEe}}
\end{figure}

In Fig.\,\ref{dGdEe} the distribution of the energy of the electron (positron) not included in the definition of the $m_{\ga\bar{e}}$ ($m_{\ga e}$) invariant mass is shown.

In a recent arXiv posting \cite{CQZ}, L.-B.~Chen, C.-F.~Qiao and R.-L.~Zhu recalculated the $m_{e\bar{e}}$ mass distributions and the corresponding widths for this Higgs boson decay mode. Their choice of cuts is rather different than ours. The closest for comparison purposes is their Cut III, which results in a width of 0.361 keV compared to our $k=0.1$ cut of 0.253 keV.

\begin{acknowledgements}
DAD was supported in part by the U.~S.~Department of Energy under grant No. DE-FG02-12ER41830. WWR was supported in part by the National Science Foundation under Grant PHY 1068020. We thank Rustem Ospanov for discussions about appropriate cuts on the photon and lepton energies.
\end{acknowledgements}


\begin{thebibliography}{98}
\bibitem{ABCDR_1} The detailed formulae of Ref.\,\cite{ABCDR} seem to be free of misprints with one exception; in the last factor of Eq. C6 the expression $+(\beta_{0+}\rightarrow\beta_{0-})$ should be $-(\beta_{0+}\rightarrow\beta_{0-})$.
\end{thebibliography}
\end{document}